# Mitigating Electrode-Induced Polarization Artifacts in Miniaturized Terahertz Detectors via a Ring-Shaped Electrode Design

*Hongjia Zhu[†], Shaojing Liu[†], Zhaolong Cao, Ximiao Wang, Runli Li, Yanlin Ke, Huanjun Chen\*, Shaozhi Deng\**

State Key Laboratory of Optoelectronic Materials and Technologies, Guangdong Province Key Laboratory of Display Material and Technology, School of Electronics and Information Technology, Sun Yat-sen University, Guangzhou 510275, China.

\*Corresponding authors: chenhj8@mail.sysu.edu.cn, stsdsz@mail.sysu.edu.cn.

[†]These authors contributed equally.

**ABSTRACT:** Terahertz (THz) polarization detection provides critical insights into material properties but faces a fundamental constraint upon miniaturization: subwavelength metallic electrodes induce strong localization and distortion of the incident field, thereby convoluting the intrinsic device response with electrode-induced artifacts. Here, we overcome this limitation with a ring-shaped electrode architecture that suppresses field perturbations across a broad bandwidth from 2.0 to 5.0 THz. The resonant frequency of the ring electrode can be flexibly detuned from the target operation frequency by adjusting its inner and outer radii, while the smooth, edge-free geometry minimizes the lightning-rod effect. These design features collectively lead to a pronounced suppression of localized THz field enhancement. Numerical simulations reveal an 8.48× reduction in the local field strength compared with conventional rod-shaped electrodes. Consistent with this, experimental measurements on graphene-based detectors exhibit a 6.95× decrease in photocurrent for the ring-shaped electrode relative to the rod-shaped configuration. Moreover, the ring geometry effectively reduces the linear polarization ratio of the photocurrent from >3 to <1.4, confirming its effectiveness in mitigating electrode-induced polarization anisotropy. Our design decouples the detection response from electrode-induced artifacts, enabling compact THz detectors that preserve intrinsic signal fidelity for high-quality polarization-resolved imaging and diagnostics.

THz waves, spanning the spectral region between microwaves and infrared radiation, possess unique penetration and finger-print properties that render them indispensable for applications such as security screening[1, 2], cosmological spectroscopy[3, 4], label-free biomedical sensing[5], and advanced communication systems[6, 7]. A fundamental requirement in modern THz detection and imaging systems is the ability to resolve the linear polarization states of THz waves[8, 9], as polarization provides critical information on material surface topology and anisotropic optical responses. Linear polarization-resolved THz detection is also essential for future 6G communication systems, which rely on multiplexed signal transmission and reception. This capability enables significant advancements in diverse fields, including target identification, biomedical diagnostics, and non-destructive testing[10-14]. However, conventional approaches to THz linear polarization detection typically depend on bulky optical components, owing to the intrinsic polarization insensitivity of most commercial THz detectors. This limitation results in increased system complexity, higher costs, and large spatial footprints.

Recently, miniaturized linear polarization-sensitive detectors have been developed by integrating artificial micro–nano structures, whose strong optical resonances enable polarization-resolved detection[15-18]. These designs provide a compact form and thereby eliminating the need for bulky optical components. However, their polarization discrimination performance in the THz frequency range remains suboptimal due to poor polarization resolved capability[19-21]. A pivotal factor underlying this limitation is the metallic electrode geometry, particularly when the footprint of the electrode pair becomes comparable to the incident THz wavelength. In this regime, the electrodes behave as an effective linearly polarized THz antenna with a narrow-band resonance. Such an antenna strongly absorbs THz radiation and induces pronounced electric-field localization around the electrode surfaces when excited at frequencies close to its resonance, and with a polarization aligned to the antenna resonant axis determined by its geometry. Consequently, the electrode pair imprints its own polarization response onto the overall detector sensitivity[22, 23], even for devices that were originally designed to probe different polarization states. Figure 1 summarizes several electrode configurations widely employed in mid-infrared and THz micro- and nanodetectors[16, 17, 24, 25]. Under THz irradiation, rod-shaped electrodes, representing the most widely employed electrode configurations, exhibit a distinct far-field scattering spectrum, with a pronounced resonance peak under linear polarization (LP) excitation, along with enhanced near-field localization along the length axis compared to orthogonal polarization (Figure 1a). More complex geometries, such as L-shaped and Y-shaped electrodes[16, 24], have been systematically explored through computational simulations, as illustrated in Figure 1b and 1c. These designs produce resonant peaks at different polarization angles, leading to complex intricate polarization-dependent THz scattering spectra (Figure 1, right panels). These results highlight that the overall device performance is governed not only by the intrinsic detecting material properties but also strongly by the electrode pair geometry that couples the detector to external circuitry. This insight underscores the urgent need to design polarization-insensitive electrode architectures, enabling THz detectors to achieve their intended polarization-resolved capabilities.

Herein, we systematically investigate, both theoretically and experimentally, how electrode geometry affects polarization-dependent response in miniaturized THz detectors. To minimize electrode-induced artifacts, we adopt a ring geometry whose symmetry yields negligible polarization response compared to conventional configurations. Targeting 2.52 THz, we optimized the inner and outer radii of the ring to shift its resonance beyond the detector bandwidth, suppressing field localization from 2.0–5.0 THz. Based on this design, we fabricated subwavelength detectors with 35 μm channel lengths by integrating either ring- or rod-shaped electrodes with monolayer graphene. The pristine graphene layer shows negligible polarization sensitivity, so the electrode pairs govern the detector response. Relative to rod electrodes, ring electrodes reduce the maximum response by a factor of 6.95 due to suppressed THz field localization. To further assess electrode effects, we patterned graphene into plasmon polariton atomic cavities (PPACs) with strong but polarization-insensitive THz absorption[26]. While rod electrodes introduced artificial polarization sensitivity, ring electrodes preserved the intrinsic characteristics. Overall, the ring design reduces the linear polarization ratio of the PPAC detector from >3 to <1.4, providing a straightforward route to compact THz detectors with reliable polarization sensitivity and minimized electrode-induced artifacts.

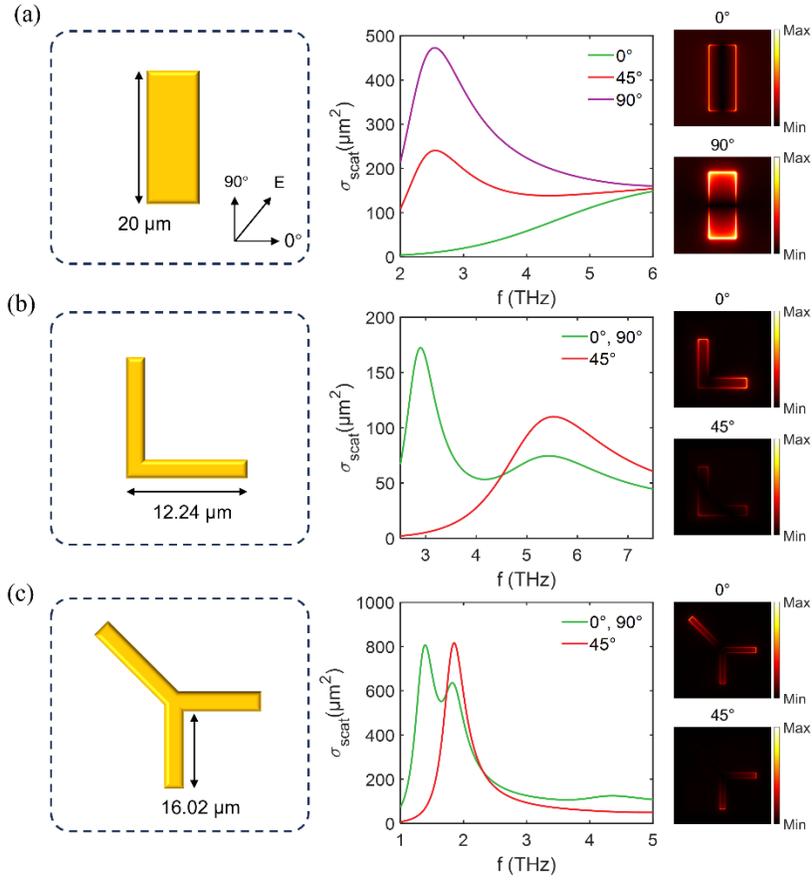

**Figure 1. Typical electrode geometries widely used in mid-infrared and THz detectors. (a–c) Rod-shaped (a), L-shaped (b), and Y-shaped (c) electrodes. The corresponding far-field scattering cross-sections and localized electric field distributions under different linearly polarized excitations are shown in the middle and right panels, respectively. The localized field distributions of the three electrodes are taken at 2.52 THz (rod-shaped electrode), 2.89 THz (L-shaped electrode), and 1.38 THz (Y-shaped electrode), respectively. The rod-shaped electrode has a length of 20 μm and a width of 8 μm. The L- and Y-shaped electrodes are adapted from Refs. 16, 24, with widths of 2.19 μm and 2.32 μm, respectively.**

**Result and discussion**

To suppress polarization artifacts arising from electrode geometry, $C_4$-symmetric designs are preferred. Square geometries satisfy this condition but generate field singularities at sharp vertices, whereas circular electrodes avoid singularities through continuous curvature yet offer limited spectral tunability since their response is governed solely by radius. Annular ring electrodes overcome these limitations by allowing independent tuning of inner and outer radii, enabling dual-parameter resonance engineering while maintaining polarization symmetry. This combination of symmetry and tunability makes the annular geometry the optimal choice for our THz detector design.

We first performed full-wave simulations to quantify the polarization-dependent responses of four electrode geometries: rod-, L-, Y-, and ring-shaped configurations. All electrodes were modeled on a silicon substrate covered with a 300-nm $SiO_2$ layer, as depicted in Figure 2b, which is the typical substrate used for fabricating photodetectors. The refractive indices for $SiO_2$ and Si in the THz regime were set at 1.955[27] and 3.42[28], respectively, with the electrode defined as a perfect electrical conductor (PEC). Figure 2a presents the far-field scattering cross-sections of these electrodes under different incident polarization angles. The electrodes exhibit pronounced polarization dependence, with the exception of the ring-shaped configuration. Specifically, in the scattering spectra, the rod- and ring-shaped electrodes exhibit a single resonance peak, whereas the L- and Y-shaped electrodes display two peaks. The mode of the rod-shaped electrode occurs around 2.5 THz, which is excited only when the polarization is aligned with the long axis, originating from the excitation of a dipole mode. The resonant modes of the L-shaped (Y-shaped) electrode appear at 2.89 (1.38) THz and 5.4 (1.81) THz, both of which can be excited under 0° or 90° polarization, whereas only a single resonance is accessible under 45° or 135° polarization. The localized electric field distributions of the L-shaped electrode at different frequencies are shown in Figure S1 and S2 (Supporting Information). For the mode at 2.89 THz (Figure S1, Supporting Information), excitation with 0° and 90° polarizations produces out-of-phase behavior between the real parts of the $E_x$ and $E_y$ components over large regions, while their imaginary parts are almost completely out-of-phase. Consequently, when the incident polarization is 45°, where both $E_x$ and $E_y$ components are equally excited, the induced currents along the two arms of the L-shaped electrode flow in opposite directions, producing out-of-phase localized fields. This destructive interference strongly suppresses the mode and leads to the disappearance of the corresponding scattering peak in the far-field spectrum. In contrast, under 90° and 180° polarization excitation, the induced currents in the two arms are aligned, and both the real and imaginary parts of $E_x$ and $E_y$ are in-phase across most localized regions. Therefore, when the polarization is at 135°, where both components are again equally excited, the two arms oscillate cooperatively, giving rise to constructive interference that enhances the mode and results in a pronounced scattering peak in the far-field response.

Conversely, the 5.4 THz mode displays the opposite phase relationship (Figure S2, Supporting Information). Under 0° and 90° polarizations, the real and imaginary parts of $E_x$ and $E_y$ oscillate largely in phase, enabling constructive interference when both components are equally excited at 45°. Under 135° polarization, however, the phase relation becomes destructive, suppressing this mode and diminishing the corresponding scattering peak. For the Y-shaped electrode, two distinct resonances emerge (Figures S3 and S4, Supporting Information). The 1.38 THz mode, dominated by the longer arm, is primarily governed by the imaginary components of the local field. Its out-of-phase response under 45° polarization leads to destructive interference, while 135° excitation restores constructive coupling and intensifies the resonance. The higher-frequency 1.81 THz mode, localized at the shorter arms, exhibits the opposite trend, i.e., enhanced at 45° and suppressed at 135°, arising from the reversal of phase relations in its field components. In contrast, the ring-shaped electrode, endowed with $C_4$ rotational symmetry, maintains uniform phase and amplitude distributions under all polarization angles, leading to an isotropic and polarization-independent scattering response.

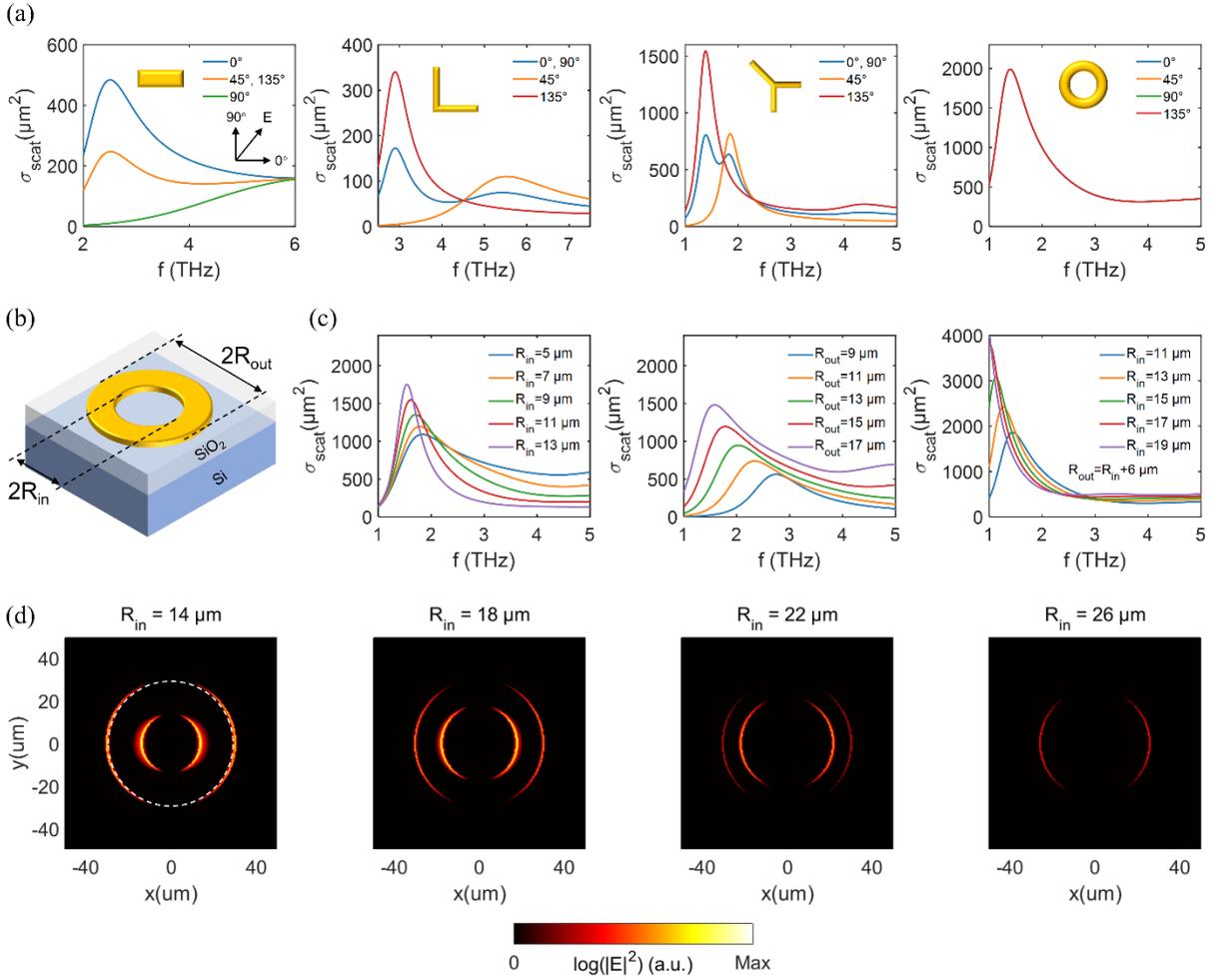

**Figure 2. Comparative THz resonance properties of ring-shaped and conventional electrodes. (a) Polarization-resolved far-field scattering cross-sections for rod-shaped (20 μm × 8 μm), L-shaped, Y-shaped (dimensions from Figure 1), and ring ($R_{out}$ = 17.5 μm, $R_{in}$ = 11.5 μm) electrodes. (b) Schematic of the ring electrode fabricated onto a silicon substrate covered with a 300-nm SiO$_2$ layer. (c) Resonance tuning via geometric parameters: (left) inner radius**

**variation at fixed $R_{out}$ = 15 μm; (middle) outer radius variation at fixed $R_{in}$ = 7 μm; (right) concurrent tuning with fixed ring width, defined as $R_{out} - R_{in}$ = 6 μm. (d) Electric field distributions at 2.52 THz for varying inner radii, with dashed line marking the electrode boundaries.**

The resonance frequency of the ring electrode can be tuned by varying its inner and outer radii. As shown in Figure 2c (left panel), increasing the inner radius at a fixed outer radius ($R_{out}$ = 15 μm) induces a gradual redshift of the resonance peak accompanied by peak narrowing. In contrast, varying $R_{out}$ at a fixed inner radius ($R_{in}$ = 7 μm) produces a larger redshift while maintaining nearly constant peak width (Figure 2c, middle panel). These results indicate that simultaneous adjustment of both radii is more effective for detuning the resonance from the operational bandwidth. Accordingly, we fixed the ring width ($R_{out} - R_{in}$) and scaled both radii proportionally, shifting the resonance below 1 THz (Figure 2c, right panel). Although this geometry preserved relatively high scattering cross-sections above 2.0 THz, further optimization, where $R_{out}$ is fixed at 30 μm and $R_{in}$ is varied from 10 to 26 μm, reduced the scattering cross-section above 2.0 THz to <500 μm$^2$ (Figure S5a, Supporting Information). This result confirms the effectiveness of configuration optimization in suppressing unwanted scattering within the operation band.

Reduction of the scattering cross-section directly results in suppression of localized electric fields, which critically influence detector performance, particularly its polarization-resolved response. Figure 2d shows the simulated field distributions of ring electrodes with varying inner radii at 2.52 THz. As the inner radius increases, the localized fields near both the inner and outer edges of the ring progressively weaken. Since the outer edge of the electrode is in direct contact with the active detection layer, field localization at this boundary is especially detrimental, as it couples strongly into the channel and introduces polarization-dependent artifacts. Our simulations reveal that when $R_{in}$ reaches 26 µm (with $R_{out}$ fixed at 30 µm), the localized field intensity at the outer edge is almost entirely suppressed, leaving only weak residual fields inside the ring cavity. Physically, this suppression originates from the reduced capacitive charging at the electrode edges as the inner radius $R_{in}$ increases, which effectively detunes the electrode resonance away from the operation band. This detuning minimizes the formation of localized hot spots with strong electric fields, thereby mitigating electrode-induced artifacts. Through this systematic evaluation of field distributions, we identify $R_{out}$ = 30 µm and $R_{in}$ = 26 µm as the optimal ring geometry for operation at 2.52 THz, ensuring minimal electrode-induced perturbations while preserving intrinsic polarization sensitivity of the active layer. It is worth noting that further suppression of the outer-edge field strength can be achieved by increasing the inner radius (Figure S5b and S5c, Supporting Information). This highlights the advantage of the two-parameter optimization strategy, which enables a balanced trade-off between minimizing electromagnetic localization and ensuring practical manufacturability in electrode design.

Figure 3 compares the field localization properties of the optimized ring-shaped electrode with those of a conventional rod-shaped electrode under different linearly polarized THz excitations. Both electrodes were connected to a PEC wire to model practical detector configurations incorporating a signal-collecting lead. The rod-shaped electrode was designed with length $L$ = 20 µm and width $W$ = 8 µm, while the connecting wire width was fixed at $d$ = 4 µm for both geometries. Excitations with polarization angles ranging from 0° to 90° were applied, with the corresponding electric field distributions shown in Figure 3c and 3d. The ring-shaped electrode maintains nearly consistent field distributions across different polarization states, with localization confined to the ring edges along the excitation direction, reflecting its inherent polarization symmetry. In contrast, the rod-shaped electrode exhibits strong field concentration at its upper and lower boundaries, particularly under 90° polarization. This pronounced localization originates from antenna-like behavior: the rod supports dipolar charge accumulation at its terminations, leading to strong field enhancement at the electrode ends where current continuity forces abrupt charge buildup. By comparison, the closed-loop geometry of the ring enables circulating surface currents without sharp terminations, thereby suppressing dipolar hot spots and minimizing polarization-dependent field localization.

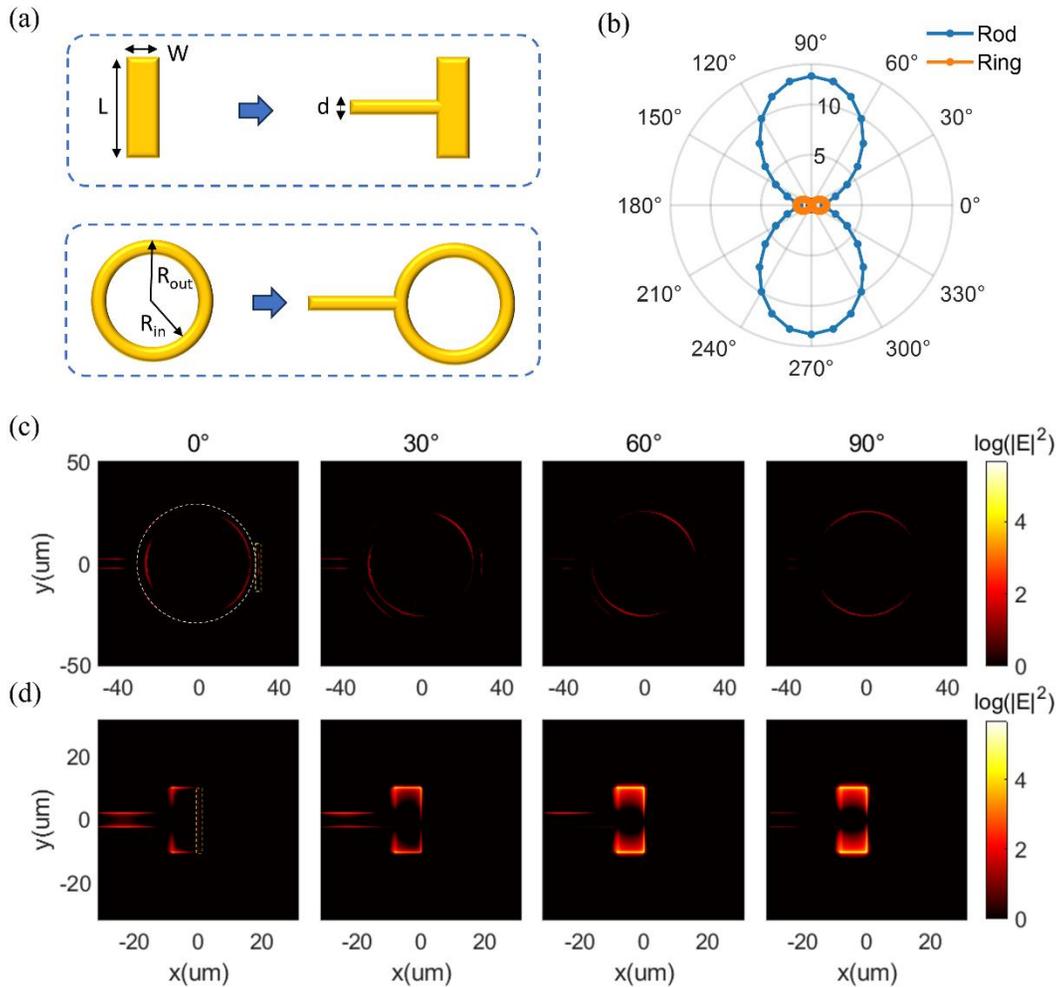

**Figure 3. Field localization characteristics of different electrodes under linearly polarized excitation at 2.52 THz. (a) Schematic of rod-shaped (top) and ring-shaped (bottom)**

**electrodes, each connected to PEC wires to emulate practical detector configurations. (b) Mean field enhancement factor (MFEF) as a function of polarization angle from 0° to 360° for both electrode types. (c, d) Polarization-dependent electric field distributions for (c) ring-shaped and (d) rod-shaped electrodes at selected polarization angles at 0°, 30°, 60°, and 90°. Dashed boxes denote the 20 μm × 2 μm electrode–active layer contact region used for MFEF evaluation. The rod electrode exhibits strong polarization-dependent end-localized fields, whereas the ring geometry maintains consistent suppressed fields across all polarization states.**

To quantify the field localization effects, we introduce the mean field enhancement factor (MFEF), *M*, defined as,

$$M = \frac{\int_C |\boldsymbol{E}|^2 dS}{\int_C |\boldsymbol{E}_0|^2 dS} \tag{1}$$

where $\boldsymbol{E}_0$ denotes the incident electric field, $\boldsymbol{E}$ the localized electric field near the electrode, and the integration domain *C* corresponds to the electrode–active layer contact region. As a representative example, *C* is indicated by the yellow dashed box, with a size of 20 μm × 2 μm, along the outer edge of the electrode in Figure 3c and 3d. The calculated MFEF values for both electrode geometries over polarization angles from 0° to 90° are summarized in Figure 3b. Under 90° polarization, the rod-shaped electrode exhibits the strongest field concentration within the contact area, consistent with its dipolar antenna-like behavior. In contrast, the ring-shaped electrode shows a markedly weaker response profile, with suppressed field intensities across all polarization states. Quantitatively, the ring-shaped geometry reduces the localized field intensity in the contact region by a factor of 8.48 compared to the rod-shaped configuration. This pronounced reduction underscores the effectiveness of ring electrode in mitigating electrode-induced field localization and preserving the intrinsic polarization response of the detector.

The suppression of electrode-induced field localization extends across a broad spectral range. Figure S6 (Supporting Information) compares near-field distributions of ring- and rod-shaped electrodes from 2.0 to 5.0 THz. The ring electrode exhibits minimal field localization under varying polarizations and frequencies (Figure S6a, Supporting Information), whereas the rod electrode shows strong edge localization under vertical polarization, with a pronounced resonance at 2.0−3.5 THz attributed to the longitudinal dipole mode (Figure S6b, Supporting Information). Even under off-resonant conditions, such as horizontal polarization, the rod electrode still exhibits markedly stronger localization than the ring. MFEF analysis (Figures S6c and S6d, Supporting Information) further underscores this contrast: the ring maintains consistently low MFEFs (<1.5 for horizontal, <0.1 for vertical), while the rod reaches values >0.95 across 2.0–5.0 THz, peaking at 18.2 at 2.5 THz under vertical polarization. Under horizontal excitation, MFEFs of the rod are lower but remain significant (1.5-4.7). To highlight the suppressed field localization of the ring-shaped electrode, we calculate its MFEF relative to the rod-shaped design: the ring electrode reaches only 0.04–1.05 of the MFEF of rods under horizontal excitation and 0.0026–0.09 under vertical polarization. These results highlight the broadband efficacy of the ring geometry in suppressing electrode-induced field localization.

We further evaluated field suppression in the ring electrode under elliptically polarized excitation. Detecting elliptically polarized THz waves is critical for Stokes-parameter detectors and advanced applications in biomedicine and nondestructive inspection[11, 14]. Localized fields from conventional electrodes compromise the accurate retrieval of polarization states and therefore introduce artifacts. To generate elliptical polarizations, a quarter-wave plate (QWP) was rotated with respect to the linear polarization direction ($\theta_Q$) of the incident wave, producing a full set of states represented on the Poincaré sphere (Figure S7a, Supporting Information). The corresponding near-field distributions (Figure S7c, Supporting Information) show that, for the ring electrode, fields localized smoothly along the edge, aligning with the major axis of the polarization ellipse, and became nearly uniform under circular polarization, with only minor deviations from the connecting wire. In contrast, the rod electrode exhibited strong end-localization, with pronounced enhancement under circular polarization. MFEF analysis further confirms these trends (Figure S7b, Supporting Information): the ring electrode generated small MFEFs at all polarization angles, reaching minima (0.78) under circular polarization ($\theta_Q = 45°$ and 135°) and maxima (1.49) under linear polarization ($\theta_Q = 0°$ and 90°). In contrast, the rod electrode exhibited the opposite trend, with MFEFs up to 4.73 times higher. These findings demonstrate the enhanced capability of the ring electrode configuration in mitigating field localization artifacts arising from electrode structures under elliptically polarized excitation, underscoring its applicability for precision polarization-resolved terahertz detection systems.

To validate the ring-shaped electrode design, we fabricated two types of detectors as testbeds, using either pristine monolayer graphene flakes or graphene PPACs as the detection layer (Figure S8, Supporting Information). Asymmetric metal contacts were applied on opposite sides of the graphene to enable electrical readout via the hot-carrier-assisted photothermoelectric (PTE) effect[29]. The operating principle of the PTE effect is illustrated in Figure S9c (Supporting Information): electrode-induced field localization generates thermal hotspots at the graphene–electrode interfaces, giving rise to a nonuniform electron temperature distribution $T(x)$ along the channel. In the presence of asymmetric Seebeck coefficients $S(x)$ at the dissimilar metal contacts, this temperature gradient produces a spatially varying potential ∇V(x), resulting in a measurable photovoltage,

$$V_{ph} = \int_0^L \nabla V(x)dx = -\int_0^L S(x)\nabla T(x)dx \qquad (2)$$

where $L$ is the channel length. The photovoltage magnitude scales directly with the temperature gradient ∇T(x), which is dictated by the degree of field localization. Stronger field localization yields higher carrier temperatures near the contacts, producing steeper gradients and larger photovoltage signals. This relationship provides a quantitative metric for comparing the suppression of localized fields between different electrode geometries through direct photovoltage measurements.

Figure 4a shows optical micrographs of fabricated detectors with pristine monolayer graphene as the detection layer integrated with different electrode configurations. Ring-shaped electrodes were designed with a fixed outer radius of 30 μm and inner radii varied from 14 to 26 μm. For comparison, rod-shaped electrodes with dimensions of 20 μm × 8 μm were fabricated, with the detection layer standardized at 35 μm × 20 μm across all devices. Simulated electric field distributions for these geometries are presented in Figure 4b, consistent with the trends observed in Figure 2d. Increasing the inner radius of the ring electrode progressively weakens the localized field intensity, confirming the effectiveness of geometric optimization in suppressing electrode-induced field localization. Electrical characterization (Figure 4c) shows linear *I*–*V* curves for all detector configurations, with resistances of 1400–2900 Ω, indicating excellent Ohmic contacts at the graphene–electrode interfaces.

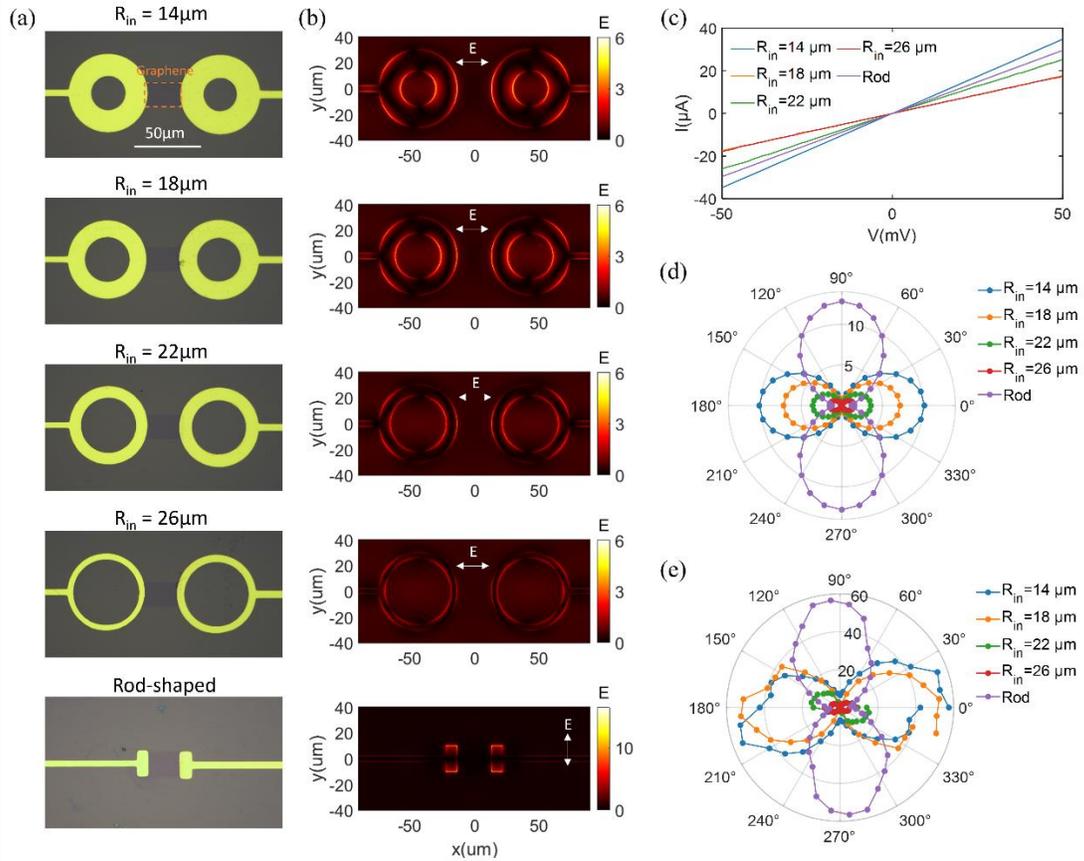

**Figure 4. Polarization-resolved THz detection of detectors consisted of different electrode configurations. (a) Optical micrographs of detectors with ring-shaped electrodes and conventional rod electrodes. The inner radii of the ring electrodes are 14–26 μm, with a $R_{out}$ fixed at 30 μm. (b) Simulated electric field distributions corresponding to the detectors shown in (a) under linearly polarized excitation. The excitation polarization is along *x*- and *y*-axis for ring electrodes and rod-shaped electrode, respectively. (c) Current–voltage characteristics showing Ohmic contact in all configurations. (d) Theoretical MFEF as a function of polarization angle, with optimized rings yielding an 8.48-fold suppression. (e) Polar plots of measured responsivity (μA/W) for rod- and ring-based detectors. The excitation frequency is 2.52 THz for all the THz detection measurements.**

To evaluate the photoresponse characteristics, we employed a far-infrared gas laser system (Edinburgh Instruments, FIRL 100) operating at 2.52 THz. The polarization dependence was characterized by rotating a half-wave plate, with the linear polarization angle defined as $\theta = 2\theta_H$, where $\theta_H$ is defined as the angle between the optical axis of the half-wave plate and the polarization direction of the incident linearly polarized light. The simulated MFEF and measured THz photocurrent responses for both electrode types are shown in Figures 4d and 4e. Under linearly polarized excitation, both configurations exhibit the characteristic "8"-shaped dipolar polar plots, confirming that the polarization sensitivity arises from electrode-induced field localization rather than graphene absorption, which is polarization-independent. For rod electrodes, the maximum response occurs at 90°, whereas ring electrodes shift the maximum to 0° and systematically suppress the polarization dependence as the inner radius increases. Quantitative analysis shows that the optimized ring design with $R_{in}$ = 26 μm reduces the simulated MFEF to 1.51, an 8.48-fold suppression compared to the rod-shaped electrode. Experimentally, the responsivity decreases from 56.73 μA/W in the rod-shaped configuration to 8.16 μA/W in the ring-shaped configuration, measured along the polarization direction that yields the maximum photovoltage (*y*-axis for rod electrode and *x*-axis for ring electrode). The photocurrent suppression reaches a factor of 6.95, closely matching the suppression observed in the MFEF. Together, these results demonstrate that ring-shaped electrodes effectively suppress artifact polarization sensitivity by mitigating field localization, providing a clear advantage over conventional rod electrodes.

To further convince that our electrode design can retrieve the intrinsic polarization properties of the detectors as desired, we replace the detection layer with a patterned graphene PPAC arrays (Figure 5a). The graphene channel features a bifurcated layer of 220 μm × 200 μm, with one half patterned as a disk-shaped PPAC array connected by conductive strips and the other half remaining unpatterned. This design leverages surface plasmon polariton resonance (SPPR)-mediated THz absorption in the PPAC arrays to generate hot electrons, producing a measurable PTE photovoltage[26]. The PPAC array was designed with a disk diameter of 5.5 μm and a periodicity of $P_x = P_y$ = 8.5 μm (Figure 5b, left panel), which supports an SPP resonance at 2.52 THz under linearly polarized excitation with various polarization angles (Figure S10, Supporting Information). The polarization response of the detector is, in principle, governed by the absorption of the disk-shaped PPAC array, which is theoretically polarization-independent (Figure 5b, right panel).

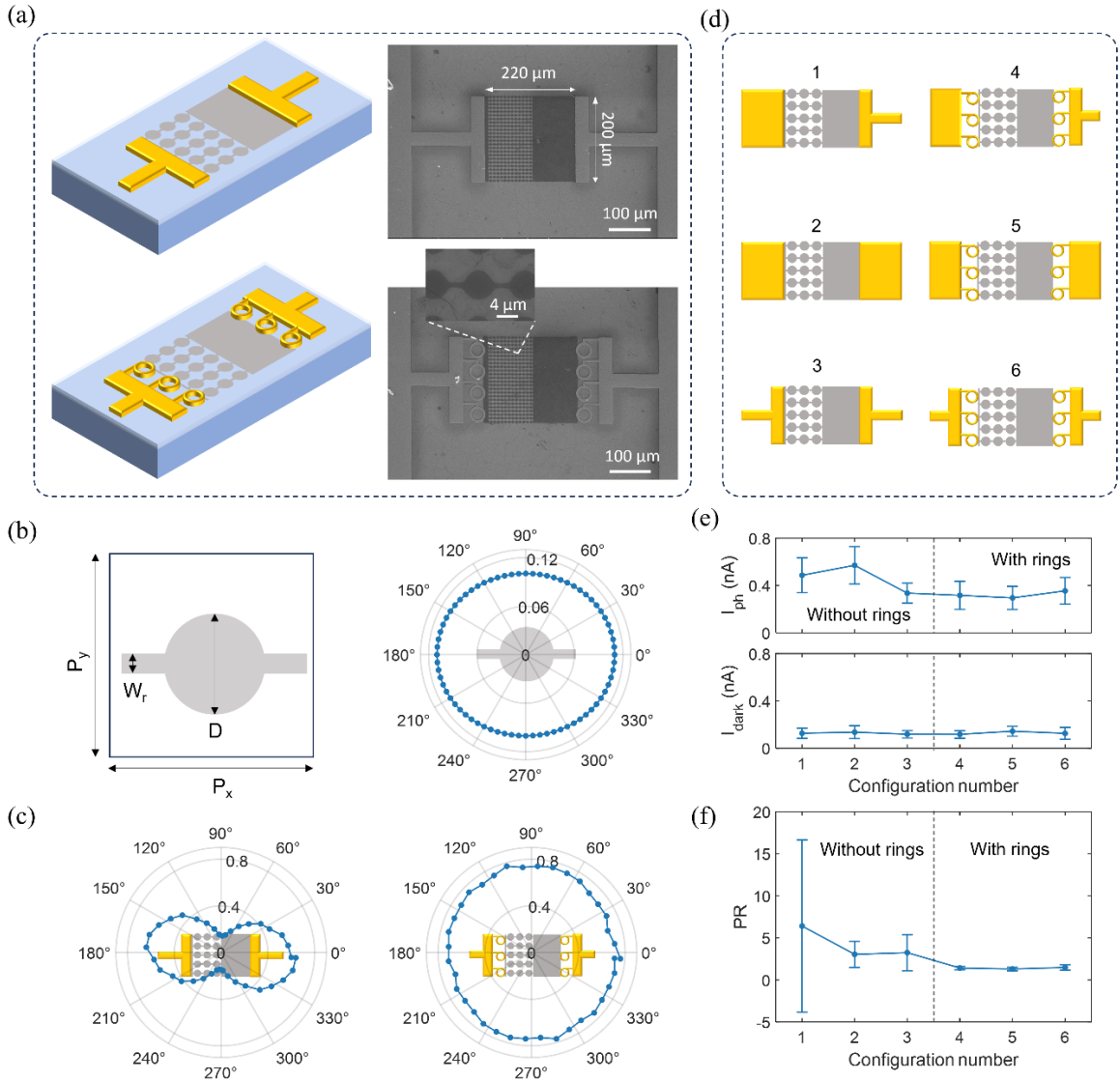

**Figure 5. Restoration of intrinsic polarization sensitivity in graphene PPAC detector using ring-shaped electrodes.** (a) Schematic and scanning electron microscope images of graphene PPAC detectors with rod-shaped (upper panel) and ring-shaped (lower panel) electrodes. The ring electrode has a dimension of $R_{out}$ = 18 μm and $R_{in}$ = 14 μm. (b) Left: Unit cell geometry of the graphene PPAC array with a periodicity of $P_x = P_y$ = 8.5 μm. The disk diameter $D$ = 5.5 μm, the connecting ribbon width $W_r$ = 0.7 μm. Right panel: Simulated absorption as a function of polarization angle at 2.52 THz, showing polarization-independent absorption. (c) Measured polarization-dependent photovoltage responsivity for detectors with rod-shaped (left) and ring-shaped (right) electrodes. (d) Evaluated electrode configurations: six designs in total, including the two shown in (a) and four additional variations (Configurations 1, 2, 4, 5). (e) Statistical comparison of photocurrent and dark current across eight devices per configuration. (f) Polarization ratio, $PR = I_{ph,max}/I_{ph,min}$, analysis showing effective suppression of electrode-induced polarization artifacts and recovery of the intrinsic, polarization-independent response of PPAC detector with ring-shaped electrodes.

To ascertain the influence of electrode geometry on the polarization sensitivity of the PPAC detector, we integrated rod- and ring-shaped electrodes on the detection channel and compared their photocurrent responses under linearly polarized excitation along different directions. The ring electrode dimensions were optimized with $R_{out}$ = 18 μm and $R_{in}$ = 14 μm to balance sufficient contact area with minimal field enhancement (MFEF < 0.3, Figure S11, Supporting Information). As shown in Figure 5c, detectors with rod-shaped electrodes exhibit pronounced polarization dependence, while those with ring-shaped electrodes display remarkable polarization insensitivity, consistent with the intrinsic absorption characteristics of the PPAC array (Figure 5b, right panel). This direct agreement between simulated absorption and experimental detector response demonstrates that conventional rod electrodes introduce strong polarization-dependent artifacts, whereas ring electrodes suppress these effects and restore the expected polarization-independent behavior.

To systematically evaluate the polarization-restoration capability of ring-shaped electrodes, we designed four additional electrode architectures (Configurations 1, 2, 4, and 5 in Figure 5d) and fabricated eight detectors per configuration for statistical characterization. Scanning electron microscopy confirmed the structural integrity of all designs (Figure S12, Supporting Information). Polarization-resolved measurements revealed marked differences in performance. While all detectors maintained comparable dark currents of ~0.1 nA, Configurations 1 and 2 exhibited substantially larger photocurrents exceeding 0.4 nA (Figure 5e), reflecting strong electrode-mediated enhancement due to the strong localized electric field. We employed the polarization ratio, defined as $PR = I_{ph,max}/I_{ph,min}$, to quantify polarization sensitivity. Conventional configurations (Configurations 1, 2, 3 in Figure 5d) yielded pronounced polarization sensitivity ($PR > 3$, Figure 5f), whereas detectors with ring-shaped electrodes (Configurations 4, 5, 6 in Figure 5d) consistently achieved near-unity polarization ratios (PR < 1.4, Figure 5f). This behavior precisely matches the simulated polarization-independent absorption of the graphene PPAC array (Figure 5b, right panel). Collectively, the statistical results in Figure 5e and 5f demonstrate that conventional electrode geometries not only exaggerate photoresponse magnitude but also impose spurious polarization dependence, while the ring-shaped design effectively suppresses these artifacts and restores the intrinsic polarization properties of the detector across all tested configurations.

## Conclusion

In conclusion, we have developed and demonstrated a ring-shaped electrode architecture tailored for monolithic THz polarization detectors. This design substantially suppresses THz field localization compared to conventional rod-shaped electrodes, particularly above 2 THz. At 2.52 THz, quantitative analysis shows an 8.48-fold reduction in maximum MFEF under linear polarization, and a 4.73-fold reduction under circular and elliptical polarizations. Across the 2.0–5.0 THz band, the MFEF ratio between rod and ring electrodes ranges from 10.8 to 377.2 under polarization excitation parallel to the length axis of the rod, underscoring the broadband suppression capability of the ring geometry. Experimental validation with graphene-based PTE detectors confirms these results, with ring-shaped electrodes yielding significantly reduced photoresponse relative to rod-shaped ones, consistent with their diminished field localization. Furthermore, when integrated into graphene PPAC detectors, ring electrodes preserve the intrinsic polarization-insensitive response of PPAC of circular shape at 2.52 THz. Statistical testing across multiple configurations highlights their ability to maintain both native polarization properties and stable photoresponse.

The performance of ring electrodes can be further refined through dimensional tuning of the inner and outer radii, enabling precise control over field localization. This architecture offers a universal solution to mitigate electrode-induced antenna effects in miniaturized THz detectors, advancing monolithic polarization-sensitive detection technologies. Beyond preserving intrinsic detector properties, the approach provides a robust platform for high-fidelity polarization detection, with potential for integration with emerging 2D materials and scalability toward array-based focal plane detection arrays.

# Method

**Device Fabrication.** The devices were fabricated on high-resistivity (>20,000 Ω·cm) silicon substrates with 300 nm thermal oxide ($SiO_2$), selected for their low terahertz absorption. Monolayer graphene grown by chemical vapor deposition (CVD) was transferred onto the substrate using a standard wet-transfer process. The graphene channels were patterned into either rectangular (Figure 4) or atomic cavity (Figure 5) configurations through maskless UV photolithography followed by oxygen plasma etching. Source and drain electrodes were defined via the same lithographic process, with subsequent thermal evaporation of metal and lift-off in acetone. Specifically, for the detectors in Figure 4, a 12 nm Cr / 70 nm Au stack and a 12 nm Bi / 70 nm Au stack were deposited on the opposing sides, respectively (Figures S9a and S9b, Supporting Information), and for the detectors in Figure 5, the same Cr/Au stacks were deposited on both sides. The Au layers ensured both reliable electrical contact and compatibility with wire bonding.

**Numerical Simulations.** All electromagnetic simulations were performed using the finite-difference time-domain method (FDTD Solutions package, Lumerical). The dielectric constants of $SiO_2$ and Si were set to 1.955 and 3.42, respectively. Electrodes were modeled as perfect electrical conductors (PEC). Graphene was implemented as a 2D material with Drude-model conductivity: $\sigma(\omega) = ie^2 E_F / (\pi\hbar^2) / (\omega + i2\Gamma)$, where the scattering rate $\Gamma = 0.001$ eV and Fermi energy $E_F = 0.32$ eV. Two distinct simulation configurations were employed: 1. For electrode characterization, a total-field scattered-field (TFSF) source with perfectly matched layer (PML) boundaries in all directions was used. 2. For atomic cavity analysis, a plane wave source with periodic boundary conditions ($x$, $y$) and PML ($z$) boundaries was implemented.

**Characterization.** Electrical measurements were conducted using a source-measure unit (Keithley 2636B). The photoresponse was characterized using a 2.52 THz far-infrared laser (Edinburgh Instruments FIRL100) delivering >50 mW of horizontally polarized output. The polarization state was controlled via a precision rotational stage-mounted half-waveplate, with power monitored by a calibrated THz power meter (National Institute of Metrology, THz-PM). The collimated beam was focused to a 2 mm spot size using a 5 cm focal length off-axis parabolic mirror. The photocurrent was amplified (FEMTO DLPCA-200) and measured via lock-in detection (Sine Scientific Instrument, OE1201) synchronized to a mechanical chopper (Stanford, SR560).

**Responsivity Calculation.** The detector responsivity was quantitatively characterized through systematic photocurrent measurements, where the generated photocurrent ($\Delta I$) was calculated from the lock-in amplifier voltage signal ($L_{lock}$) using the relation[30]: $\Delta I = \pi\sqrt{2}L/2G$, with $G$ representing the transimpedance gain of the low-noise current preamplifier. The current responsivity ($R_I$) was then determined by normalizing the photocurrent to the incident power density[30], expressed as $R_I = \Delta I S_t / (P_t S_a)$, where $S_t$ = 3.14 mm² denotes the terahertz beam spot area (2 mm diameter), $P_t$ = 50−90 mW is the calibrated optical power measured in situ using a THz power meter, and $S_a$ represents the effective detector area - specifically 0.0007 mm² for the rectangular channel devices (Figure 4) and 0.044 mm² for the atomic cavity configurations (Figure 5). Polarization-resolved measurements were conducted over 0°-180°, with data symmetrically extended to 360° via the relation $R_I(\theta_E + 180°) = R_I(\theta_E)$ to visualize the full angular dependence, where $\theta_E$ denotes the incident electric field's polarization angle.

## ASSOCIATED CONTENT

**Supporting Information**. The Supporting Information is available free of charge on the ACS Publications website.

Field distributions of L-shaped and Y-shaped electrodes;

Simulated scattering spectra and field distributions of ring-shaped electrodes;

Frequency-dependent field localization characteristics of electrodes;

Field localization characteristics of electrodes under elliptical polarization;

Electrode designs of graphene detectors;

Operational principle of graphene terahertz photodetectors;

Simulated polarization-resolved absorption of the graphene PPAC array;

Optimization of field localization characteristics for ring-shaped electrodes;

Polarization response characterization of multiple detector configurations.

## AUTHOR INFORMATION

**Corresponding Authors**

**Huanjun Chen** – *School of Electronics and Information Technology, Sun Yat-sen University, Guangzhou 510275, China*

E-mail: chenhj8@mail.sysu.edu.cn

**Shaozhi Deng** – *School of Electronics and Information Technology, Sun Yat-sen University, Guangzhou 510275, China*

E-mail: stsdsz@mail.sysu.edu.cn

**Authors**

**Hongjia Zhu** – *School of Electronics and Information Technology, Sun Yat-sen University, Guangzhou 510275, China*

**Shaojing Liu** – *School of Electronics and Information Technology, Sun Yat-sen University, Guangzhou 510275, China*

**Zhaolong Cao** – *School of Electronics and Information Technology, Sun Yat-sen University, Guangzhou 510275, China*

**Ximiao Wang** – *School of Electronics and Information Technology, Sun Yat-sen University, Guangzhou 510275, China*

**Runli Li** – *School of Electronics and Information Technology, Sun Yat-sen University, Guangzhou 510275, China*

**Yanlin Ke** – *School of Electronics and Information Technology, Sun Yat-sen University, Guangzhou 510275, China*

**Notes**

The authors declare no competing financial interest.

**Author Contributions**

H.C. and S.D. conceived and initiated the study. Sample fabrication was performed by H.Z., S.L., X.W., and R.L.. H.Z., S.L., and Y.K. conducted the experiments and analyzed the data. H.Z. and Z.C. performed the numerical simulations. H.C. and S.D. coordinated and supervised the work and discussed and interpreted the results. H.Z., H.C., and S.D. co-wrote the manuscript with the input of all other coauthors. †H.Z. and S.L. contributed equally. All authors have given approval to the final version of the manuscript.

**Funding Sources**

National Natural Science Foundation of China (grant no. 92463308). National Key Basic Research Program of China (grant nos. 2024YFA1208500 and 2024YFA1208501).

ABBREVIATIONS

MFEF, mean field enhancement factor; PPAC, plasmon polariton atomic cavity; PEC, perfect electrical conductor; THz, terahertz; LP, linear polarization; QWP, quarter-wave plate; PTE, photothermoelectric; SPPR, surface plasmon polariton resonance; PR, polarization ratio; CVD, chemical vapor deposition; FDTD, finite-difference time-domain method; TFSF, total-field scattered-field; PML, perfectly matched layer.